\documentclass[
prl,
twocolumn,
superscriptaddress,
a4paper,
showpacs,
showkeys,
floatfix
]{revtex4}
\usepackage[final]{graphicx}
\usepackage{amsmath}
\usepackage{amssymb}

\begin{document}

\title{Soccer: is scoring goals a predictable Poissonian process?}

\author{A. Heuer}
\affiliation{\frenchspacing Westf\"alische Wilhelms Universit\"at M\"unster, Institut f\"ur physikalische Chemie, Corrensstr.\ 30, 48149 M\"unster, Germany}
\author{C. M\"uller}
\affiliation{\frenchspacing Westf\"alische Wilhelms Universit\"at M\"unster, Institut f\"ur physikalische Chemie, Corrensstr.\ 30, 48149 M\"unster, Germany}
\affiliation{\frenchspacing Westf\"alische Wilhelms Universit\"at M\"unster, Institut f\"ur organische Chemie, Corrensstr.\ 40, 48149 M\"unster, Germany}
\author{O. Rubner}
\affiliation{\frenchspacing Westf\"alische Wilhelms Universit\"at M\"unster, Institut f\"ur physikalische Chemie, Corrensstr.\ 30, 48149 M\"unster, Germany}


\begin{abstract}
The non-scientific event of a soccer match is analysed on a strictly scientific level.  The analysis is based on the recently introduced concept of a team fitness
(Eur. Phys. J. B 67, 445, 2009) and requires the use of finite-size scaling.
A uniquely defined function is derived which quantitatively predicts
the expected average outcome of a soccer match in terms of the fitness of both teams. It is checked whether temporary fitness fluctuations of a team hamper
the predictability of a soccer match.
 To a very good approximation
scoring goals during a match can be characterized as independent Poissonian processes with pre-determined expectation values. Minor correlations give
rise to an increase of the number of draws. The non-Poissonian overall goal distribution is just a consequence of the fitness distribution among different teams.
The limits of predictability of soccer matches are quantified.
Our model-free classification of the underlying ingredients determining the outcome of soccer matches can be generalized
to different types of sports events.
\end{abstract}

\maketitle


In recent years different approaches, originating from the physics
community, have shed new light on sports events, e.g. by studying
the behavior of spectators \cite{laola}, by elucidating the
statistical vs. systematic features behind league tables
\cite{ben1,ben2,buch}, by studying the temporal sequence of ball
movements \cite{mendes} or using extreme value statistics
\cite{Suter,Greenhough02} known, e.g., from finance analysis
\cite{Stanley}. For the specific case of soccer matches different
models have been introduced on phenomenological grounds
\cite{Lee97,Dixon97,Dixon98,Rue00,Koning00,Dobson03}. However,
very basic questions related, e.g., to the relevance of systematic
vs. statistical contributions or the temporal fitness evolution
are still open. It is known that the distribution of  soccer goals
is broader than a Poissonian distribution
\cite{Greenhough02,janke,janke09}. This observation has been
attributed to the presence of self-affirmative effects during a
soccer match\cite{janke,janke09},  {i.e. an increased
probability to score a goal depending on the number of goals
already scored by that team}.

In this work we introduce a general model-free approach which allows us to elucidate the outcome of
sports events. Combining strict mathematical
reasoning, appropriate finite-size scaling and comparison with actual data
all ingredients of this framework can be quantified for the specific example of soccer.  A unique relation can be derived to calculate
the expected outcome of a soccer match and three hierarchical levels of statistical influence can be identified.
As one application we show that the skewness of the distribution of soccer goals \cite{Greenhough02,janke,janke09} can be fully
related to fitness variations among different teams and does not require the presence of self-affirmative effects.

As data basis we take all matches in the German Bundesliga (www.bundesliga-statistik.de) between seasons 1987/88 and 2007/08 except for the year 1991/92 (in that year the league contained 20 teams). Every team plays 34 matches per season.
Earlier seasons are not taken into account because the underlying statistical properties (in particular number of goals per match) are somewhat different.

Conceptually, our analysis relies on recent observations in describing
soccer leagues \cite{HR09}: (i) The home advantage is characterized by a team-independent but season-dependent increase of the home team goal
difference $c_{home}>0$. (ii) An appropriate observable to characterize the fitness of a team $i$ in a given season is the average goal
difference (normalized per match) $\Delta G_i(N)$, i.e. the difference of
the goals scored and conceded during $N$ matches. In particular it contains more information about the team fitness than, e.g., the number of points.

\begin{figure}[ht]
\includegraphics[width=0.95\linewidth]{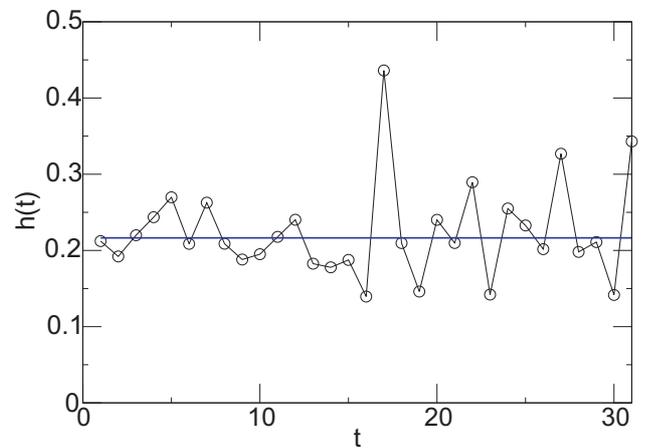}
\caption{ The correlation function $h(t)$. The average value of $h(t)$ is included (excluding the value
for $t=17$) yielding approx. $0.22$ \cite{HR09}.}
\label{fig.1}
\end{figure}

Straightforward information about the team behavior during a season can be extracted from correlating its match
results from different match days. Formally, this is expressed by
 the correlation function $h(t) = \langle \Delta g_{ij}(t_0) \Delta g_{ik}(t_0+t) \rangle$. Here
$\Delta g_{ij} := g_i - g_j$ denotes the goal difference of a match of team $i$
vs. team $j$ with the final result $g_i:g_j$. $j$ and $k$ are the opponents of
team $i$ at match days $t_0$ and $t_0 + t$. The home-away asymmetry
can be taken into account by the transformation $\Delta g_{ij} \rightarrow \Delta  g_{ij} \mp c_{home}$
where the sign depends on whether team $i$ plays at home or away.
The resulting function $h(t)$ is shown in Fig.1. Apart from the data point
for $t=17$  one observes a time-independent positive plateau value. The absolute value of this constant corresponds
to the variance $\sigma_{\Delta G}^2$ of $\Delta G_i$ and is thus a measure for the fitness variation in a league \cite{HR09}.
Furthermore, the lack of any decay shows that the fitness of a team is constant during the whole season.
This result is fully consistent with the finite-size scaling analysis in
Ref.\cite{HR09} where additionally the fitness change between two seasons was
quantified. The exception for $t=17$
just reflects the fact that team $i$ is playing against the same
team at days $t_0$ and $t_0 + 17$, yielding additional
correlations between the outcome of both matches (see also below).

As an immediate consequence, the limit of $\Delta G_i(N)$
for large $N$, corresponding to the true fitness $\Delta G_i$, is well-defined. A consistent estimator for $\Delta G_i$, based on the information
from a finite number of matches, reads
\begin{equation}
\label{eqgest}
\Delta G_i = a_N \Delta G_i(N).
\end{equation}
with $a_N \approx 1/[1+3/(N \sigma^2_{\Delta G})]$ \cite{HR09} .
For large $N$ the factor $a_N$ approaches unity and the estimation
becomes error-free, i.e. $\Delta G_i(N)  \rightarrow \Delta G_i$.
For $N=33$ one has $a_N = 0.71$ and the variance of the estimation error is
given by $ \sigma_{e,N}^2 = (N/3+1/\sigma_{\Delta G}^2)^{-1} \approx 0.06$ \cite{HR09}.  This statistical framework
is known as regression toward the mean \cite{Stigler}.
Analogously, introducing
$\Sigma G_i(N)$ as the average sum of goals scored and conceded by team $i$ in $N$ matches
its long-time limit is estimated via $\Sigma G_i -\lambda = b_N (\Sigma G_i(N)-\lambda)$ where $\lambda$ is the average number of goals
per match in the respective season. Using $\sigma^2_{\Sigma G} \approx 0.035$ one correspondingly obtains $b_{N=33} = 0.28$ \cite{HR09}.

 {Our key goal is to find a sound characterization of the match result when team $i$ is playing vs. team $j$, i.e. $\Delta g_{ij}$
or even $g_i$ and $g_j$ individually. The final outcome $\Delta g_{ij}$ has three conceptually different and uncorrelated contributions
\begin{equation}
\Delta g_{ij} = q_{ij} + f_{ij} + r_{ij}.
\end{equation}
Averaging over all matches one can define the respective variances $\sigma_q^2, \sigma_f^2$ and $\sigma_r^2$.
(1) $q_{ij}$ expresses the average outcome which can be expected based on knowledge of the team fitness values $\Delta G_i$ and $\Delta G_j$, respectively.
Conceptually this can be determined by averaging over all matches when teams with these fitness values play against each other. The task is
to determine the dependence of  $q_{ij} \equiv  q(\Delta G_i, \Delta G_j) $ on  $\Delta G_i$ and $\Delta G_j$.
(2) For a specific match, however, the outcome can be {\it systematically}
influenced by different factors beyond the general fitness values using the variable $f_{ij}$ with a mean of zero:
(a) External effects such as  several players which are injured or tired,
weather conditions (helping one team more than the other), or red cards. As a consequence the effective fitness of a team relevant for this match may differ from the
estimation $\Delta G_{i}$ (or $\Delta G_{j}$). (b) Intra-match effects depending on the actual course of a match. One example is the suggested presence
of self-affirmative effects, i.e. an increased probability to score a goal (equivalently an increased fitness) depending
on the number of goals already scored by that team \cite{janke,janke09}.  Naturally, $f_{ij}$ is much harder to predict if possible at all. Here we restrict
ourselves to the estimation of its relevance via determination of $\sigma_f^2$.
(3) Finally, one has to understand the emergence of the actual goal distribution based on expectation values as expressed by the random
variable $r_{ij}$ with average zero. This
problem is similar to the physical problem when a decay rate (here corresponding to $q_{ij} + f_{ij}$) has to be translated into the actual
number of decay processes. }

{\it Determination of $q_{ij}$}:
$q_{ij}$ has to fulfill the two basic conditions (taking into account the home advantage):
$q_{ij}-c_{home} = - (q_{ji}-c_{home})$ (symmetry condition) and
 $\langle q_{ij}\rangle_j -c_{home} = \Delta G_i$  (consistency condition) where the average is over
all teams $j \ne i$ (in the second condition a minor correction due to the finite number of teams in a league is neglected).
The most general dependence on $\Delta G_{i,j}$ up to third order, which is compatible
with both conditions, is given by
\begin{equation}
\label{eqg1}
 q_{ij} = c_{home} + (\Delta
G_i - \Delta G_j) \cdot [ 1 - c_3(\sigma^2_{\Delta G} + \Delta G_i
\Delta G_j ) ].
\end{equation}
Qualitatively, the $c_3$-term takes into account the possible
effect that in case of very different team strengths (e.g. $\Delta
G_i \gg 0$ and $\Delta G_j \ll 0$) the expected goal difference is even more
pronounced ($c_3 > 0$: too much respect of the weaker team) or
reduced ($c_3 < 0$: tendency of presumption of the better team).
On a phenomenological level this effect is already considered in
the model of, e.g., Ref.\cite{Rue00}. The task is to determine the
adjustable parameter $c_3$ from comparison with actual data. We
first rewrite Eq.\ref{eqg1} as $q_{ij} - (\Delta G_i - \Delta G_j)
- c_{home} = - c_3(\Delta G_i - \Delta G_j)(\sigma^2_{\Delta G} +
\Delta G_i \Delta G_j )$. In case that $\Delta G_{i,j}$ is known
this would correspond to a straightforward regression problem of
$\Delta g_{ij} - (\Delta G_i - \Delta G_j) - c_{home}$ vs.
$-(\Delta G_i - \Delta G_j)(\sigma^2_{\Delta G} + \Delta G_i
\Delta G_j )$. An optimum estimation of the fitness values for a
specific match via Eq.\ref{eqgest} is based on $\Delta
G_{i,j}(N)$, calculated from the remaining $N=33$ matches of both
teams in that season . Of course, the resulting value of
$c_3(N=33)$ is still hampered by finite-size effects, in analogy
to the regression towards the mean. This problem can be solved by
estimating $c_3(N)$ for different values of $N$ and subsequent
extrapolation to infinite $N$ in an $1/N$-representation. Then our estimation
of $c_3$ is not hampered by the uncertainty in the determination of $\Delta
G_{i,j}$.  For a
fixed $N \le 30$  the regression analysis is based on 50 different
choices of $\Delta G_{i,j}(N)$ by choosing different subsets of
$N$ matches to improve the statistics. The result is shown in
Fig.2. The estimated error results from performing this analysis
individually for each season. Due to the strong correlations for
different $N$-values the final error is much larger than suggested
by the fluctuations among different data points. The data are
compatible with $c_3 = 0$.  {Thus, we have shown that the simple
choice
\begin{equation}
\label{eqfinal}
q_{ij} = \Delta G_i - \Delta G_j + c_{home}
\end{equation}
is the uniquely defined relation (neglecting irrelevant terms of
5th order) to characterize the average outcome of a soccer match.}
In practice the right side can be estimated via Eq.\ref{eqgest}.
 {This result implies that $h(t) = \langle (\Delta G_i -
\Delta G_j)(\Delta G_i - \Delta G_k) \rangle = \sigma_{\Delta G}^2
+ \langle \Delta G_j \Delta G_k \rangle$, i.e. $h(t \ne 17) =
\sigma_{\Delta G}^2$ and $h(t = 17) = 2\sigma_{\Delta G}^2$. This
agrees very well with the data.} Furthermore, the variance of the
$q_{ij}$ distribution, i.e. $\sigma_q^2$, is by definition given
by $2 \sigma_{\Delta G}^2 \approx 0.44$.

\begin{figure}[ht]
\includegraphics[width=0.95\linewidth]{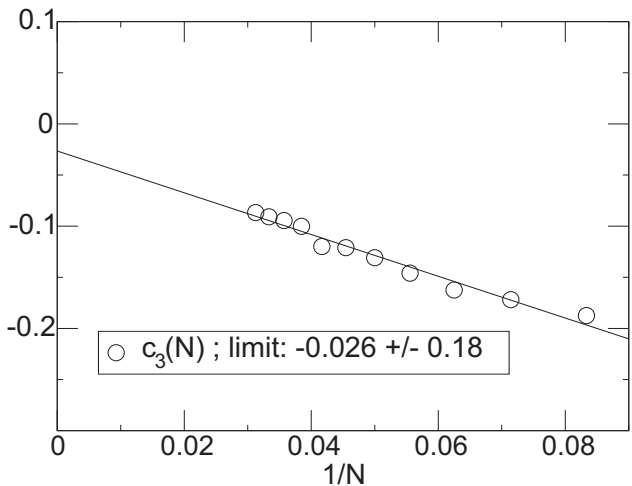}
\caption{Determination of $c_3$ by finite-size
scaling.}
\label{fig.2}
\end{figure}

{\it Determination of $\sigma_f^2$}: This above analysis does not contain any information about the
match-specific fitness relative to $\Delta G_i - \Delta G_j$. For example $f_{ij} > 0$
during a specific match implies that team $i$ plays better than
expected from $q_{ij}$. The conceptual problem
is to disentangle the possible influence of these fitness fluctuations
from the random aspects of a soccer match. The key idea is based on the observation
that, e.g., for $f_{ij} > 0$ team $i$ will play better than expected in both
the first and the second half of the match. In contrast, the random features of a match
do not show this correlation.  For the identification of $\sigma_f^2$ one defines $A =
\langle ((\Delta g^{(1)}_{ij}/b_1-c_{home})\cdot ((\Delta g^{(2)}_{ij}/b_2-c_{home})\rangle_{ij}$
where $\Delta g^{(1),(2)}_{ij}$ is the goal
difference in the first and second half in the specific match,
respectively and  $b_{1,2}$ the fraction of goals scored during the first
and the second half, respectively ($b_1 =
0.45; b_2 = 0.55$). Based on Eq.\ref{eqfinal} one has $\sigma_f^2 = A - 2\sigma_{\Delta G}^2$. Actually,
to improve the statistics we have additionally used different partitions of the match (e.g. first and third
quarter vs. second and fourth quarter).
Numerical evaluation yields $\sigma_f^2 = -0.04 \pm 0.06$ where the error bar is estimated
from individual averaging over the different seasons. Thus one obtains in particular
$\sigma_f^2 \ll \sigma^2_q$ which renders match-specific fitness fluctuations irrelevant.
Actually, as shown in \cite{HR09}, one can observe a tendency that teams which have lost 4 times
in a row tend to play worse in the near future than expected by their fitness. Strictly speaking
these strikes indeed reflect minor temporary  fitness variations. However, the number of strikes
is very small (less than 10 per season) and, furthermore, mostly of statistical nature. The same holds
for red cards which naturally influence the fitness but fortunately are quite rate.  Thus, these extreme
events are interesting in their own right but are not relevant for the overall statistical description.
The negative value of $\sigma_f^2$ points towards anti-correlations between both partitions of the match.
A possible reason is the observed tendency towards a draw, as outlined below.

\begin{figure}[ht]
\includegraphics[width=0.95\linewidth]{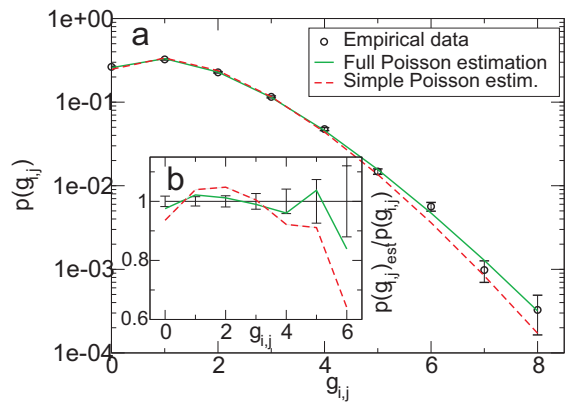}
\caption{({\bf a})Distribution of goals per team and match and the Poisson prediction if the different fitness values
are taken into account (solid line). Furthermore a Poisson estimation is included where only the home-away asymmetry is included
(broken line). The quality of the predicted distribution is highlighted in ({\bf b}) where the ratio of the estimated and the actual probability is shown.}
\label{fig.3}
\end{figure}

{\it Determination of $r_{ij}$}: The actual number of
goals $g_{i,j}$ per team and match is shown in Fig.3. The error
bars are estimated based on binomial statistics. As discussed
before the distribution is significantly broader than a Poisson
distribution, even if separately taken for the home and away goals
\cite{Greenhough02,janke,janke09}. Here we show that this
distribution can be generated by assuming that scoring goals are
independent Poissonian processes. We proceed in two steps. First,
we use Eq.\ref{eqfinal} to estimate the average goal difference
for a specific match with fitness values estimated from the
remaining 33 matches of each team. Second, we supplement
Eq.\ref{eqfinal} by the corresponding estimator for the sum of the
goals $g_i + g_j$ given by $ \Sigma G_i + \Sigma G_j-
\lambda$. Together with Eq.\ref{eqfinal} this allows us to
calculate the expected number of goals for both teams individually.
 Third, we generate for both teams a Poissonian
distribution based on the corresponding expectation values. The
resulting distribution is also shown in Fig.1 and perfectly agrees
with the actual data up to  8 (!) goals. In contrast, if the
distribution of fitness values is not taken into account
significant deviations are present.  { Two conclusions can be drawn. First,
scoring goals is a highly random process. Second, the good
agreement again reflects the fact that $\sigma_f^2$ is small because
otherwise an additionally broadening of the actual data would be expected.
Thus there is no indication of a possible influence of
self-affirmative effects during a soccer match  \cite{janke,janke09}}.
Because of the underlying  Poissonian process the value of $\sigma_r^2$ is just given by the average number of
goals per match $(\approx 3)$.

\begin{figure}[ht]
\includegraphics[width=0.95\linewidth]{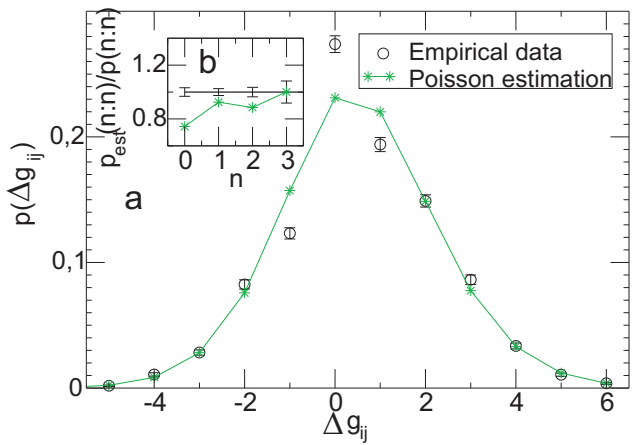}
\caption{({\bf a}) The probability distribution of the goal difference per match together
with its estimation based on independent Poisson processes of both teams. In
({\bf b}) it is shown for different scores how the ratio of the estimated and the
actual number of draws differ from unity.}
\label{fig.4}
\end{figure}

As already discussed in literature the number of draws is
somewhat larger than expected on the basis of independent Poisson distributions; see, e.g., Refs. \cite{Dixon97,Rue00}.
As an application of the present results
we quantify this statement. In Fig.4 we compare the calculated distribution of $\Delta g_{ij}$
with the actual values. The agreement is very good except for
$\Delta g_{ij} = -1,0,1$. Thus, the simple picture of independent goals of the home and the away
team is slightly invalidated. The larger number of draws is balanced by a
reduction of the number of matches with exactly one goal
difference. More specifically, we have calculated the relative
increase of draws for the different results. The main effect is due to the
strong increase of more than 20\% of the 0:0 draws. Note that the present analysis has
already taken into account the fitness distribution for the estimation of this number. Starting from 3:3 the
simple picture of independent home and away goals holds again.

The three major contributions to the final soccer result
display a clear hierarchy, i.e. $\sigma_r^2 : \sigma_q^2:
\sigma_f^2  \approx 10^2:10^1:10^0$.
$\sigma_f^2$, albeit well defined and quantifiable,  can be
neglected  for two reasons. First, it is small as compared to the
fitness variation among different teams. Second, the uncertainty
in the prediction of $q_{ij}$ is, even at the end of the season,
significantly larger (variance of the uncertainty: $2 \cdot
\sigma_{e,N=33}^2 = 0.12$, see above).  {Thus, the limit of predictability of a soccer
match is, beyond the random effects, mainly related to the uncertainty in the
fitness determination rather than to match specific effects}. Thus, the hypothesis of a
strictly constant team fitness during a season, even on a
single-match level cannot be refuted even for a data set
comprising more than 20 years. In disagreement with this
observation soccer reports in media often stress that a team
played particularly good or bad. Our results suggest that there
exists a strong tendency to relate the assessment too much to the
final result thereby ignoring the large amount of random aspects
of a match.

In summary, apart from the minor correlations with
respect to the number of draws soccer is a surprisingly simple
match in statistical terms. Neglecting the minor differences
between a Poissonian and binomial distribution and the slight
tendency towards a draw a soccer match is equivalent to two teams
throwing a dice. The number 6 means goal and the number of
attempts of both teams is fixed already at the beginning of the
match, reflecting their respective fitness in that season.

More generally speaking, our approach may serve as a general
framework to classify different types of sports in a
three-dimensional parameter space, expressed by  $\sigma_r^2,
\sigma_q^2, \sigma_f^2$. This set of numbers, e.g., determines the
degree of competitiveness  \cite{ben2}. For example for matches
between just two persons (e.g. tennis) one would expect that
fitness fluctuations ($\sigma_f^2$) play a much a bigger role and
that for sports events with many goals or points  (e.g.
basketball) the random effects ($\sigma_r^2$) are much less
pronounced, i.e. it is more likely that the stronger team indeed wins.
Hopefully, the present work stimulates activities to
characterize different types of sports along these lines.

We greatly acknowledge helpful discussions with B. Strauss, M. Trede, and M. Tolan about this topic.

\end{document}